\documentstyle[11pt,newpasp,twoside, psfig]{article}
\newcommand{\Msun}{\ifmmode {M_{\odot}}\else${M_{\odot}}$\fi~}
\newcommand{\Rsun}{\ifmmode {R_{\odot}}\else${R_{\odot}}$\fi}
\def\Lx{L$_X$~}
  
\def\Edot{$\dot{E}$~}
\def\Chandra{{\it Chandra}~}
\def\HST{{\it HST}~}
\def\ergss{ergs s$^{-1}$}

\def\edcomment#1{\iffalse\marginpar{\raggedright\sl#1\/}\else\relax\fi}
\reversemarginpar

\begin{document}
\title{X-ray and Optical Studies of Millisecond Pulsars in 47 Tucanae}
 \author{C. O. Heinke, J. E. Grindlay, P. D. Edmonds}
\affil{Harvard-Smithsonian Center for Astrophysics, 60 Garden Street,
 Cambridge, MA  02138, USA}
\author{F. Camilo}
\affil{Columbia Astrophysics Laboratory, Columbia University, 550 West
 120th Street, New York, NY 10027, USA}

\begin{abstract}
Our \Chandra X-ray observation of the globular cluster 47 Tuc clearly
detected most of the 16 MSPs with precise radio positions, and
indicates probable X-ray emission from the remainder.  The MSPs are
soft (BB $kT\sim$0.2-0.3 keV) and faint (\Lx$\sim$ few $10^{30}$
\ergss), and generally consistent with thermal emission from small
polar caps.  An additional 40 soft X-ray sources are consistent with
the known MSPs in X-ray colors, luminosity, and radial distribution
within the cluster (and thus mass).  We note that these MSPs display a
flatter \Lx to \Edot relation than pulsars and MSPs in the field,
consistent with polar cap heating models for younger MSPs and may 
suggest the surface magnetic field has been modified by repeated
accretion episodes to include multipole components.  Correlating HST
images, radio timing positions, and the \Chandra dataset has allowed
optical searches for MSP binary companions.  The MSP 47 Tuc-U is
coincident with a blue star exhibiting sinusoidal variations that
agree in period and phase with the heated face of the WD companion.
Another blue variable star (and X-ray source) agrees in period and
phase with the companion to 47 Tuc-W (which lacks an accurate timing
position); this companion is probably a main sequence star.
\end{abstract}

\index{binaries: eclipsing}
\index{pulsars}
\index{globular clusters: individual (NGC 104)}
\index{stars: neutron}

X-ray and optical studies of the millisecond pulsars (MSPs) in 47 Tuc
  (Freire et al. 2001a; Lorimer, these proceedings) allow 
  us to probe the physics of the polar caps, the evolutionary state of the
  binary systems in which many reside, and the complete population of
  millisecond pulsars, many of which have not yet been discovered in the
  radio.  \Chandra is the only
  X-ray telescope capable of resolving the faint MSPs
  from the quiescent LMXBs, cataclysmic variables (CVs), and active
  (flaring) main-sequence binaries 
  in the crowded central regions of globular clusters, and 
  {\it Hubble Space Telescope} (\HST) archival data allows deep
  searches for the  intrinsically faint white dwarf (WD) binary
  companions expected for the  MSPs.  
 
\section{\Chandra X-ray Studies}

   Using the source-detection
tool WAVDETECT, we have identified 103 sources in the central 
2\arcmin$\times$2.5\arcmin region containing the 16 MSPs with known
timing positions (Grindlay et al. 2001, GHE01).  Ten identified X-ray
sources were 
identified with timing positions of 12 MSPs
(including the unresolved pairs F \& S, G \& I); 
aligning the radio and X-ray frames revealed probable emission
from the locations of the remaining four (not formally detected due to
crowding with brighter sources, or for 47 Tuc-C, low exposure on a
chip gap).   The MSPs show a remarkable similarity in luminosities;
 all are clearly identified with 0.5-2.5 keV luminosities in the range
$10^{30.0-30.6}$ \ergss except for two unresolved pairs and 47 Tuc-C,
 also consistent with this range.  No correlation is seen
between the X-ray and radio luminosities.

The MSPs are clearly ``softer'' than the majority of the other X-ray
sources.  Grindlay et al. 2002 (GCH02)
plotted two hardness ratios for the identified MSPs (Fig. 1a),
along with predicted ratios for simple spectral shapes (using the
column density to 47 Tuc).  Most of the MSPs, with the
notable exception of 47 Tuc-J, are consistent with 0.2 to 0.3 keV
blackbody emission and inconsistent with a simple powerlaw of photon
index 1 to 2.  This implies that most of the MSPs exhibit
predominantly thermal emission from part of their polar caps; the
temperature and emitting radius 
($\sim0.2$ km for a BB, 0.8 km for H-atmosphere model) are consistent
with the higher-temperature thermal component identified by Zavlin et
al. (2002) in the nearby MSP J0437-4715.

\begin{figure}

\psfig{file=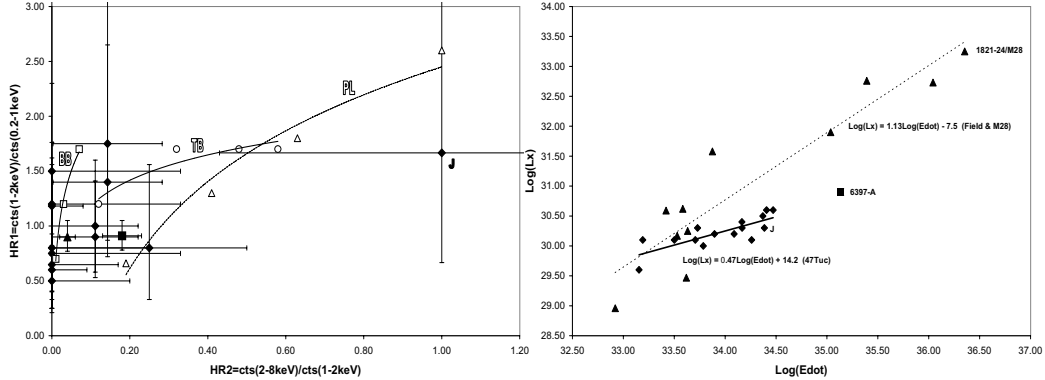}

\vspace{0.5cm}

\caption[heinkec1_1.eps]{Left: X-ray color-color diagram of the MSP
emission (diamonds), vs. blackbody (BB, open squares; kT=0.2, 0.25,
0.3 keV), thermal bremss 
(TB, open circles; kT=1, 2, 3, 6 keV), and powerlaw (PL, open
triangles; photon index=3, 2, 1.5, 1) 
PIMMS models, parameters increasing to upper right except for PL.
Weighted mean colors are shown for all MSPs (square) and all but 47
Tuc-J (triangle).  Right: \Lx vs. \Edot for the MSPs of 47 Tuc
(diamonds) and field MSPs (triangles).  The best-fit slopes are also plotted.
Both from GCH02. \label{Figure 1}} 
\end{figure}

GCH02 compare the relation of the 0.5-2.5 keV X-ray luminosities of the 47 Tuc
MSPs to their 
rotational energy loss \Edot (calculated for these globular cluster
MSPs using a three-dimensional model of the cluster gas and potential,
Freire et al. 2001b) to that for field MSPs, following Becker \&
Tr\"umper (1999).   
 In contrast to the field MSPs, for which $L_X\propto\dot{E}^{1.13\pm0.15}$
(consistent with emission from pulsars generally),
we find that the 47 Tuc MSPs show $L_X\propto\dot{E}^{0.5\pm0.2}$ (see
Fig. 1b).
This flatter dependence of \Lx on \Edot is predicted by the models of
Harding \& Muslimov (2002) for polar cap heating.  However, the MSPs
retain this dependence to larger characteristic ages than predicted by
the model, leading GCH02 to propose that the MSPs have had their
surface magnetic fields altered toward multipole configurations
through repeated accretion episodes.

\begin{figure}
\psfig{file=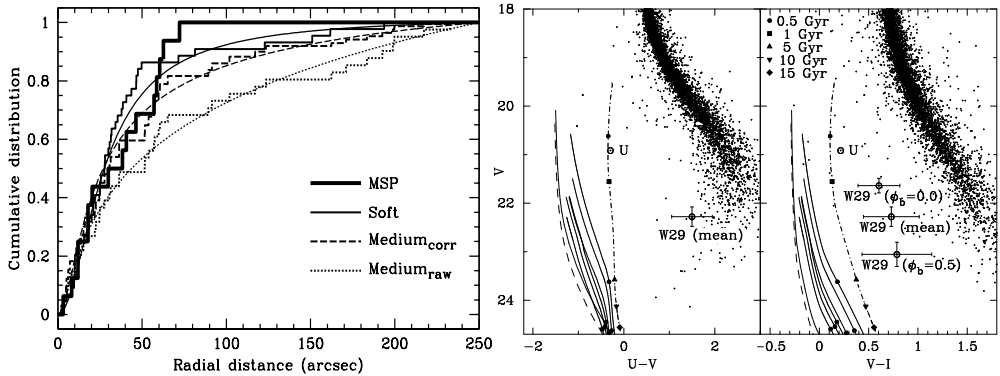}

\vspace*{0.5cm}

\caption[heinkec2_2.eps]{Left: Radial distributions of the MSPs, soft
sources, and medium sources (before and after correcting for
background AGN), along with best-fit generalized King model
predictions for each group (GCH02).  Right: \HST CMDs (based on PSF-fitting)
showing the relative position of W29 (47 Tuc-W) and 47 Tuc-U.  Cooling
WD model predictions are shown; 47
Tuc-U is consistent with a Serenelli et al. (2001) 0.169 \Msun He WD
cooling track, while W29 seems to be a heated main-sequence star (EGC02).
\label{Figure 2}}
\end{figure}

GCH02 also investigated the larger populations of \Chandra sources out
to a radius of 4'.  The MSP radial distribution is 
consistent with that of the soft sources, according to a
Kolmogorov-Smirnov test, requiring no abrupt truncation of their
spatial 
distribution (Fig. 2a).  The  soft sources are also consistent with 
a ``generalized King model'' (Lugger, Cohn \& Grindlay 1995) profile,
with a probable mass range of 0.9-1.4 M$_{\odot}$, and 
include CVs and active binaries as well as unidentified MSPs.  
GCH02 estimate that, if the 
radio and X-ray luminosities are uncorrelated, the number of MSPs in
47 Tuc lies in the range 35-90. 

\section{\HST Optical Studies}

  Using the detection of six
CVs and active binaries in the \Chandra and \HST data, we have fixed
the \HST and radio coordinate systems, allowing  the detection of the
first MSP companion in a 
globular cluster, 47 Tuc U (Edmonds et al. 2001).  A blue star was
identified at the radio position (see fig. 2b), and low-amplitude
sinusoidal variations were seen in the V-band data, with a period matching
the radio orbital period.  In addition, the phase of maximum brightness
corresponds to the phase of maximum visibility of the side of the WD
expected to be heated by the MSP wind.  The position of 47 Tuc-U in
the CMD indicates that the companion is indeed a helium WD of mass
$\sim0.17$ \Msun.

Another MSP optical counterpart has been identified
{\it without} a radio timing position for the MSP (Edmonds et al. 2002,
EGC02).  The \Chandra X-ray 
source W29 (GHE01) was identified with a blue star showing
large-amplitude (60-70\%) sinusoidal variability.  Phase-connecting
\HST data from several programs, including recent ACS calibration
data, produced an accurate period, 0.132944(1) days, matching the
radio orbital  period of 47 Tuc-W within 0.5 s.  The phase of
optical maximum also matches the phase of maximum visibility of the
heated side of the companion within 1.2 minutes (0.006 in
phase).  Camilo et
al. (2000) noted that 47 Tuc-W has a short period (3.2 hours), but a much
higher mass than other short-period systems ($\sim0.15$ \Msun), and
eclipses in the radio.  Its CMD location
suggests that the companion is a main-sequence star, significantly
heated by the MSP wind but not Roche-lobe filling.  This implies that
the companion is probably the result of an exchange encounter
(common in dense cluster cores), and not
the original companion.  

Additional \Chandra and simultaneous \HST data will be collected by
October 2002 (PI J. Grindlay), allowing deeper studies of the known MSPs and
identification of other MSPs not yet detected in the radio.

\acknowledgments

This work was supported in part by Chandra grant GO0-1034A.  We thank
R. Gilliland, H. Cohn, and P. Lugger.


\begin{references}

\reference Becker, W. \& Tr\"umper, J. 1999, A\&A 341, 803
\reference Camilo, F., Lorimer, D.~R., Freire, P., Lyne, A.~G., \&
Manchester, R.~N. 2000, ApJ 535, 975
\reference Edmonds, P. D., Gilliland,
R.~L., Heinke, C.~O., Grindlay, J.~E., \& Camilo, F. 2001, ApJ 557, L57
\reference Edmonds, P. D., Gilliland,
R.~L., Camilo, F., Heinke, C.~O., \& Grindlay, J.~E.  2002, ApJ (in
press, astro-ph/0207426; EGC02)
\reference Freire, P.~C. et al. 2001a, MNRAS 326, 901
\reference Freire, P.~C. et al. 2001b, ApJ 557, L105
\reference {Grindlay}, J.~E., {Heinke}, C.~O., {Edmonds}, P.~D., \& {Murray},
S.~S. 2001a,  Science 292, 2290 (GHE01)
\reference {Grindlay}, J.~E., {Camilo}, F., {Heinke}, C.~O.,
{Edmonds}, P.~D.,  {Cohn}, H., \& {Lugger},
P. 2002, ApJ (in press, astro-ph/0208280; GCH02)
\reference Harding, A.~K. \& Muslimov, A.~G. 2002, ApJ 568, 862
\reference Lugger, P.~M., Cohn, H.~N., \& Grindlay, J.~E. 1995 ApJ
439, 191
\reference Serenelli, A.~M., Althaus, L.~G., Rohrmann, R.~D., \&
Benvenuto, O.~G. 2001, MNRAS 325, 607
\reference {Zavlin}, V.~E. et al. 2002, ApJ 569, 894


\end{references}
\end{document}